\documentstyle[preprint,aps,floats]{revtex}
\input psfig.tex
\begin{document}

\draft

\title{A Preposterous Universe\footnote{{\sc Perspective: Astronomy},
                               published in Science 299, 1333-1334 (2003) 
                               (28 February 2003). }}
\author{Alejandro Gangui\footnote{The author is at {\sc IAFE}, the
        Institute for Astronomy and Space Physics (CONICET), and at the Physics
        Department, University of Buenos Aires, Ciudad Universitaria,
        1428 Buenos Aires, Argentina. E-mail: gangui@df.uba.ar}}

\maketitle
\hspace{0.2in}

For centuries, astronomers have wondered about the mechanisms behind the formation of galaxies and
large-scale structures. In the second half of the 20th century, cosmologists realized that the
witness of it all was a hot bath of light, now cooled to a few degrees Celsius above absolute zero,
which is the afterglow of the big bang. The sky-pervading cosmic microwave background (CMB)
radiation\cite{ref1} was released just before the epoch when matter begun getting structured. Some
ten years ago, the tiny variations discovered in its effective temperature\cite{ref2} told us about
the size of the primordial seeds leading to the nascent galaxies, after eons of gravitational
evolution.

Now, another piece of evidence fits in the right place and shows us the way these primordial seeds
were actually moving some 400,000 years after the bang, when the CMB decoupled from non-relativistic
matter. With a radio telescope at the South Pole, scientists from the DASI
collaboration\cite{ref3,ref4} have recently measured the minute level of orientation, or
polarization, that these microwaves received when they emerged from the seething plasma -- a signal
that only the peculiar dynamics of the seeds present at that epoch can generate\cite{ref5,ref6}.

Most light around us is unpolarized. Its many individual waves oscillate in different planes as it
propagates. But unpolarized light becomes polarized whenever it is scattered or reflected, as in
sunglasses or in the surface of a lake. In these cases, most of the intensity of the scattered light
is concentrated in one plane along the line of propagation, resulting in linearly polarized light.

Early on, when the universe was hot enough, matter was ionized and the free electron density was so
high that photons could not propagate freely without colliding with electrons. But the universe was
expanding and the ambient temperature decreasing, and so the energetic collisions became less
frequent. The relatively low-energy photons that ensued could not destroy the increasing number of
neutral particles (essentially hydrogen and helium) that began to combine. Cosmologists refer to this
period as ``recombination'', and soon afterward the CMB was released free, making its last scattering
upon matter. According to theory, it is at this precise time, nearly 14 billion years ago, that the
CMB became polarized. 

CMB polarization was first discussed thirty-five years ago by Martin Rees\cite{ref7}. However, there
was no evidence of its existence until the DASI detection late last year. Polarization is an
important probe for cosmological models and for the more recent history of our nearby universe. It
arises from the interaction of the cosmic background radiation with free electrons; hence, CMB
polarization can only be produced at the time of its last scattering\footnote{With the formation of
the first stars and quasars, and the subsequent UV radiation emited by these primitive sources, the
hydrogen can re-ionize. As a consequence, the CMB will scatter again upon ionized matter and will
also modify its polarization, albeit on a different angular scale.}. Unlike temperature fluctuations,
polarization is largely unaffected by inhomogeneities in the growing distribution of matter after
recombination.

To understand how the CMB becomes polarized, two points should be clear. First, the oscillating
electric field of the incoming radiation will push the electron to also oscillate; the latter can
then be seen as an electric dipole, and dipole radiation emits preferentially perpendicularly to the
direction of oscillation (see Figure 1). Second, after interaction with the electron, the resulting
radiation field will be polarized with the same orientation as the incident electromagnetic
wave. These rules will help us understand why the CMB should be linearly polarized.

\begin{figure}[t]
\centerline{\psfig{file=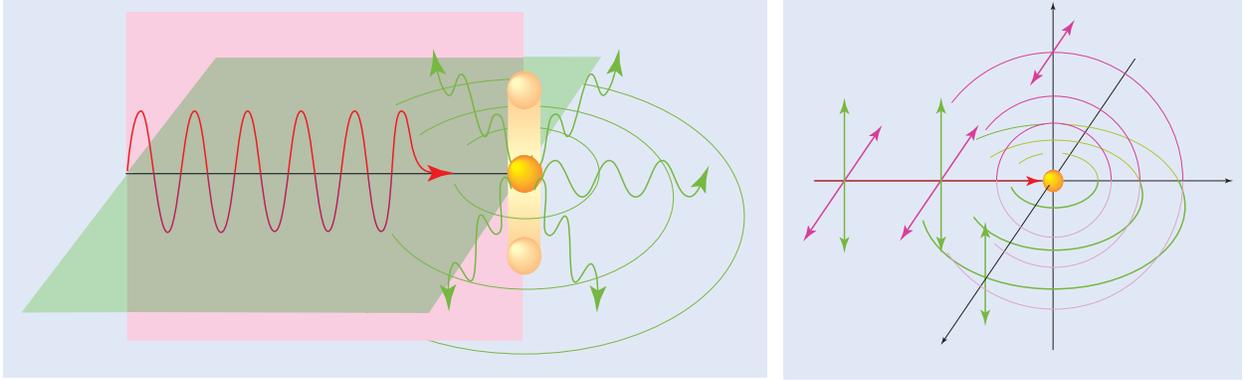,width=6.5in}}
\vspace{0.1in}
\caption{{\bf Playing tricks with light.} Left panel: An electromagnetic linearly polarized wave (in
red) oscillates in a given plane (in pink). Reaching an electron (orange ball), the wave induces the
electron to also oscillate, making it emit radiation (in green). This resulting wave is concentrated
essentially in the (green) plane orthogonal to the movement of the electron and is polarized like the
incident wave. Right panel: Non-polarized light can be decomposed into the sum of two linearly
polarized waves: one along the line of sight (in pink), the other along a perpendicular direction (in
green). Scattered radiation due to the first wave is contained in the plane orthogonal to the line of
sight and cannot be detected. Only the second component (in green) reaches the observer and is
polarized like the incident wave.}
\end{figure}

Before the recombination epoch, the radiation field was unpolarized. In unpolarized light, the
transverse electric field can be decomposed into the two orthogonal directions (pink and green arrows
in Figure 1) to the line of propagation. The electric field component along the vertical direction
(green arrow) will make the electron oscillate also vertically. Hence, the dipolar radiation will be
maximal over the horizontal plane. Looking from the side, we can easily detect this
component. However, the incident component oriented along our line of sight (pink arrow) will not
reach us, just because this component causes the electron to also oscillate along our line of sight,
emitting the scattered dipole radiation on the orthogonal plane. From our position then we cannot
perceive this radiation. Thus, it is as if only the vertical component of the incoming electric field
has caused the radiation we perceive. Recalling now the second rule mentioned above, the resulting
radiation reaching the observer should have the same polarization as the incident vertical component
(green arrow). In conclusion, the observer only receives one part of the incident radiation, and this
fraction is linearly polarized.
 
But so far, we are leaving aside the fact that, in the real case, the target electron will be hit by
radiation waves coming from all possible directions, independently. Thus, to convey the total effect,
we need to consider all these contributions, and sum them up. Each of these waves will be scattered
to the observer in the form of linearly polarized radiation, but each one with a different
orientation. If the incoming radiation were completely isotropic in intensity, then radiation coming,
say, from the left would provide the polarization state that is missing from that coming from above
the electron, leaving the net outgoing radiation unpolarized: just from symmetry arguments, in a
spherically symmetric configuration no direction is privileged.

But the CMB is not perfectly isotropic. It has a tiny ``quadrupole anisotropy'', first discovered by
experiments onboard of the COBE satellite\cite{ref2}. Hence, from any point of view, the orthogonal
contributions to the final polarization will be different, leaving a net linear polarization in the
scattered radiation. Theorists believe that this is how the CMB polarization detected by DASI, and
recently confirmed with data from the Wilkinson Microwave Anisotropy Probe (WMAP)
satellite\cite{ref8}, arose. 

There is one last point to emphasize. Before recombination, ionized matter, electrons and radiation
formed a single fluid. In this fluid, inertia was provided by massive nucleons whilst the pressure
was that of radiation. And this fluid supported sound waves: the gravitational clumping tendency of
the effective mass in the perturbations was resisted by the restoring radiation pressure, and
therefore gravity-driven acoustic oscillations in both the fluid density and local velocity
appeared. Now, the polarization field responds to the local quadrupole moment at recombination, and
this quadrupole is mainly due to the Doppler shifts induced by the velocity field of the
plasma\cite{ref9}. That is why we know with certainty that the CMB polarization shows the
uncontaminated dynamics of the primordial seeds at recombination.

However, the polarized fraction of the temperature anisotropy is small, since only those photons that
scattered at the very last instants of the decoupling process possess a sufficiently large quadrupole
moment that is not washed out in subsequent interactions. The reason is that the scattering that
generates polarization also suppresses the quadrupole from which polarization arises. The result is
that the polarized fraction of the temperature anisotropy is no more than a 10\%. Since the
temperature anisotropies are at the $10^{-5}$ level, the polarization signal was expected to be at
the $10^{-6}$ level, just a few microkelvin -- a phenomenal experimental challenge.

The level of polarization detected by the DASI and WMAP collaborations was
compared\cite{ref5,ref6,ref8} with predictions of one of the currently most favored theoretical
models, the so-called concordance model. This is the model that best fits the bulk of current
astrophysical observations, and contains 5\% of ordinary matter, 22\% of dark matter, and the rest in
dark energy in the form of Einstein's cosmological constant (or something even more bizarre): a truly
preposterous universe. Comparing their results with the prediction of the concordance model,
scientists claim that polarization has been detected with a 95\% confidence level.

As more independent experiments confirm and extend these findings, the CMB polarization will become
the next gold mine of cosmology. A new window on the physics of the early universe is opening wide in
front of us. Through it, we expect to obtain key information on the fundamental parameters of
cosmology -- and perhaps even the actual mechanism behind the formation of large-scale structures in
our universe.

\begin{figure}[htbp]
\centerline{\psfig{file=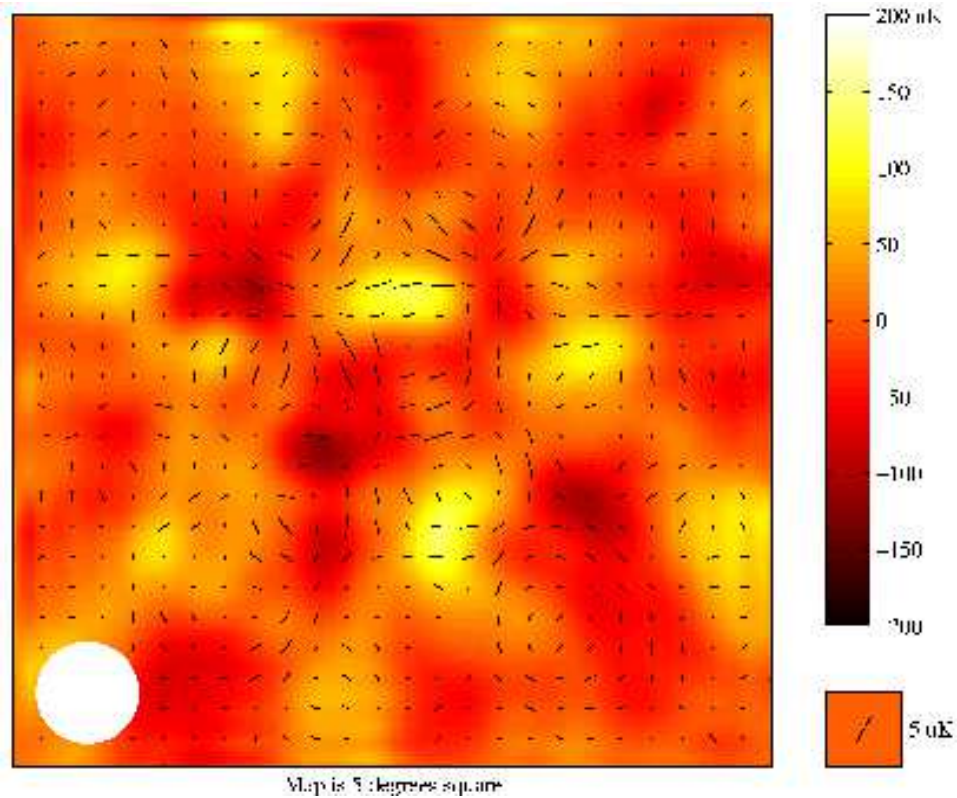,width=5in}}
\vspace{0.1in}
\caption{{\bf A mesure of motion.}  Intensity and polarization of the cosmic microwave background
(CMB) radiation measured with the DASI telescope. The small temperature variations of the CMB are
shown in false color, with yellow indicating hot and red cold. The polarization at each spot in the
image is shown by a black line. The length of the line shows the strength of the polarization; its
orientation indicates the direction in which the radiation is polarized. The size of the white spot
(lower left) approximates the angular resolution of the observations.}
\end{figure}

\vspace{0.5in}

\begin{quote}
{\em ``Basta que sea irracional un solo hombre para que otros lo sean } \par
{\em y para que lo sea el universo''. } \par
{\em La historia universal abunda en confirmaciones de ese temor. } \par
$~~~~~~~~~~~~$ --Jorge L. Borges (1944), Pr\'ologo a {\em Bartleby} de Herman Melville.
\end{quote}

\end{document}